\documentclass[12pt]{article}
\usepackage{amsmath,amsfonts,amssymb}
\usepackage{float}
\usepackage[letterpaper, margin=.75in]{geometry}
\usepackage{graphicx,caption,subcaption}% Include figure files
\usepackage{wrapfig}
\usepackage{bm}% bold math
\usepackage{authblk}
\usepackage{amsfonts}
\usepackage{hyperref}
\hypersetup{
    colorlinks=true,
    linkcolor=black,
    filecolor=magenta,      
    urlcolor=blue,
    citecolor=black,
}
 
\usepackage[capitalise]{cleveref}
\usepackage{xcolor}

\newcommand{\te}[1]{\mbox{\boldmath$ #1 $}} 

\def\bx{\te{x}}

\def\bI{\te{I}}
\def\bU{\te{U}}

\def\bR{\te{R}}

\def\bF{\te{F}}

\def\bn{\te{n}}
\def\bL{\te{L}}

\def\b1{\te{1}}

\def\bq{\te{q}}

\def\bOmega{\te{\Omega}}
\def\bepsilon{\te{\epsilon}}

\usepackage[backend=biber,style=nature,sorting=none,date=year,doi=false,isbn=false,url=false,eprint=false]{biblatex}

\makeatletter

\begin{document}
\title{Ray Optics for Gliders}

\author[1,*]{Tyler D. Ross}
\author[2]{Dino Osmanovi{\'{c}}}
\author[3]{John F. Brady}
\author[1]{Paul W. K. Rothemund}
\affil[1]{Department of Computing and Mathematical Sciences, California Institute of Technology, Pasadena, CA 91125, USA.}
\affil[2]{Center for the Physics of Living Systems, Department of Physics, Massachusetts Institute of Technology, Cambridge, Massachusetts 02139, USA}
\affil[3]{Divisions of Chemistry \& Chemical Engineering and Engineering \& Applied Science, California Institute of Technology, Pasadena, CA 91125, USA}
\affil[*]{correspondence to: tross@caltech.edu}

\date{}

\maketitle

\begin{abstract}

%The field of ray optics, used to control the trajectories of photons, provides us inspiration: a single principle, Snell's law, yields an intuitive framework for engineering a broad range of devices, from microscopes to cameras and telescopes.
%In ray optics, Snell's law provides a single principle for controlling the path of light

%One attractive approach is for trajectory to be determined by interaction of %the particle with local environmental cues. 

Control of self-propelled particles is central to the development
of many microrobotic technologies, from dynamically reconfigurable materials to advanced lab-on-a-chip systems. However, there are few physical principles by which particle trajectories can be specified and can be used to generate a wide range of behaviors. Within the field of ray optics, a single principle for controlling the trajectory of light---Snell's law---yields an intuitive framework for engineering a broad range of devices, from microscopes to cameras and telescopes. Here we show that the motion of self-propelled particles gliding across a resistance discontinuity is governed by a variant of Snell's law, and develop a corresponding ray optics for gliders. Just as the ratio of refractive indexes sets the path of a light ray, the ratio of resistance coefficients is shown to determine the trajectories of gliders. The magnitude of refraction depends on the glider's shape, in particular its aspect ratio, which serves as an analog to the wavelength of light. This enables the demixing of a polymorphic, many-shaped, beam of gliders into distinct monomorphic, single-shaped, beams through a friction prism. In turn, beams of monomorphic gliders can be focused by spherical and gradient friction lenses. Alternatively, the critical angle for total internal reflection can be used to create shape-selective glider traps. Overall our work suggests that furthering the analogy between light and microscopic gliders will result in a wide range of new devices for sorting, concentrating, and analyzing self-propelled particles.

    % Snell's law, which encompasses both refraction and total internal reflection, provides the foundation for ray optics and all lens-based instruments, from microscopes to telescopes. Refraction results when light crosses the interface between media of different refractive index, the dimensionless number that captures how much a medium retards the propagation of light. In this work, we show that the motion of self-propelled particles moving across a resistance discontinuity is governed by an analogous Snell's law, allowing for glider ray optics. We derive a variant of Snell's law for gliders moving across regions of different frictions. Just as the ratio of refractive indexes sets the path of a light ray, the ratio of resistance coefficients is shown to determine the trajectories of gliders. We find that the magnitude of refraction depends on the glider's shape, specifically the aspect ratio, as analogous to the wavelength of light. This enables the demixing of a polymorphic, many-shaped, beam of gliders into distinct monomorphic, single-shaped, beams through a friction prism. In turn, beams of monomorphic gliders can be focused by spherical and gradient friction lenses. Completing the analogy, we show that the shape-dependence of the total internal reflection critical angle can be used to create glider traps. Such analogies to ray optics suggest an expansive set of new devices for sorting, concentrating, and analyzing microscopic gliders is possible.
\end{abstract}

\newpage

\section*{Introduction}
\vspace*{-0.10in} 

Biology provides a compelling existence proof that complex micro- and nanorobotics are possible \cite{Chowdhury2005,Lauga2009,Chou2011,MolecularRobots2014,ActiveMatterNeedlemanZvonomir2017,Fisher2018BioinspiredMicrorobots}. Immune cells swarm to the site of an infection \cite{ImmuneSwarm2022} and chase bacteria over many hundreds of microns. At a smaller scale, motor proteins traffic lipid droplets, vesicular containers of neurotransmitters, and even whole mitochondria from one end of the cell to the other along protein filament tracks \cite{MicrotubuleTransportReview2017}. Thus we imagine that a mature microrobotics will enable us to deliver drugs precisely to where they are needed within the body \cite{Peng2017,Li2017,Luo2018}, and to create advanced labs-on-a-chip in which analyte molecules are moved and sorted autonomously \cite{TurberfieldRobot2011, QianCargoSort2017}, without the need for pumps and valves \cite{Snchez2014,Ebbens2016}. Current approaches to microrobotics vary widely in degree of autonomy and control---from reprogramming the sensors and circuits of biological cells so that they independently seek out a novel chemical \cite{ArtificialChemotaxisLim2014}, to precision steering of artificial metal spirals with a global magnetic field under computer control \cite{metalhelix2010}. Here, our interest is in simple self-propelled particles, which (1) move autonomously but can perform no complex logic themselves, and (2) have trajectories which are determined by purely mechanical interactions with their local environment. Such particles are relatively easy to fabricate and need no special instrumentation for control. Self-propelled particles are often classified by their modes of microscale motion, e.g. gliding or swimming \cite{henrichsen1972glidingdef}. In particular we study gliders, rigid particles which propel themselves via interaction with a solid surface or matrix \cite{epifanio2014dissipativeshocksgliding,SHROUT2015244}. Our aim is to provide a general theory and framework for controlling the trajectories of autonomous gliders that is flexible enough to accomplish multiple microrobotic tasks.

\begin{figure}[H]
    \centering
    \includegraphics[width=\textwidth]{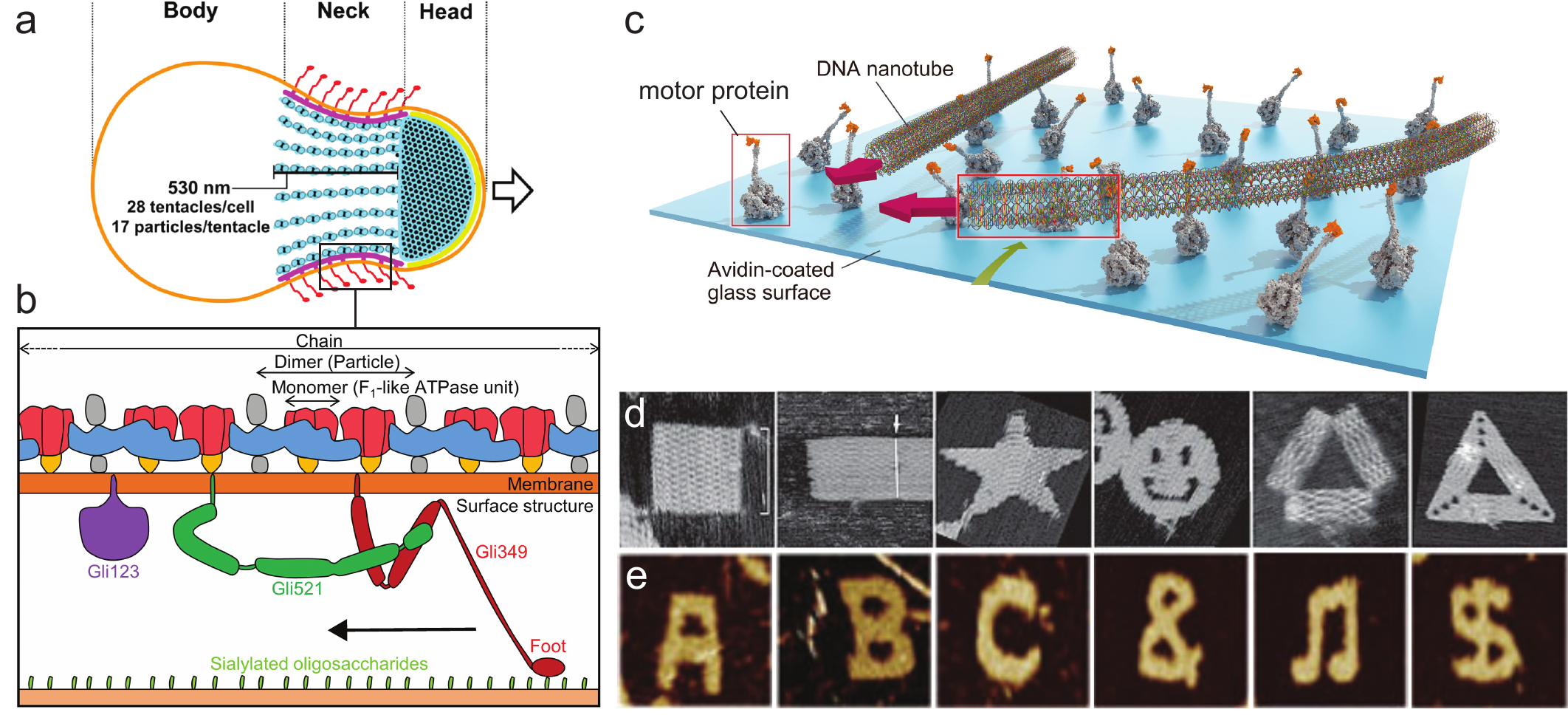}
    \caption{Experimental glider systems and DNA nanostructures provide  motivation for this work. \textbf{a--b}, Diagram of \emph{M. mobile} (a) and a zoom of its gliding machinery (b) adapted from \cite{MiyataMycoplasma2019} and \cite{ToyonagaMBio2021}, respectively. \textbf{c}, Diagram of DNA nanotube / protein motor gliding system adapted from \cite{IbusukiScience2022}. For both \emph{M. mobile} and DNA nanotube glider systems, binding between the glider and a coating of molecules on the surface create a viscous-like drag force \cite{tawada1991protein,bormuth2009protein,erbas2012viscous}. \textbf{d--e}, Atomic force microscopy images of $\sim$100~nm diameter DNA origami (d) and single stranded DNA tiles (e) shapes adapted from \cite{rothemund2006folding} and \cite{SST2012} show that any desired glider shape should be constructible. Standard DNA nanotechnology techniques can be used to decorate any shape with the same DNA recognition sequences as those used in (c) to connect motors with the nanotubes so that, in principle, any shape can be made to glide. }
    \label{fig:gliderbkg}
\end{figure}

So far, topographically and chemically-patterned surfaces have been used to explicitly confine gliders to a chosen trajectory. In this way, gliding protein filaments have been constrained to follow paths \cite{clemmenshess2003chemicaltopography} or navigate junctions \cite{clemmenshess2004junctions} and the microorganism \emph{Mycoplasma mobile} has been placed on an enclosed track to drive a microrotor \cite{HiratsukaPNAS2006}. However, these approaches have limited ability to route and sort gliders---gliders of different shapes cannot easily be distinguished, for example. Two lines of recent work suggest another approach. First, experiments have found that more complicated self-propelled particles, such as flexible crawling cells and swimmers, refract and scatter at adhesion or viscosity discontinuities \cite{lopez2021dynamics, Kantsler1187algae,Coppola2021,theriot2011,BoundariesSteerJanusSpheres2015, AbaurreaVelasco_FlexocyteRefraction2019}. Second, theoretical studies of swimmers in viscosity gradients have predicted viscotaxis, the degree to which swimmer trajectories bend toward or away from high viscosity, as a function of swimmer symmetry and swimming mechanism \cite{Liebchen2018,elfring2019gradientswimmers,stehnach2021viscophobic}. Together these findings suggest that self-propelled particles might be treated similar to light moving through a medium of varying refractive index, with a resistivity replacing refractive index. Deriving a simple theory for resistive refraction in these systems, however, is challenging as even the direction of viscotaxis (towards or away from high viscosity) varies based on the details of the specific system. Fortunately, for gliders, the effects of friction can be isolated from particle flexibility, propulsion mechanism, and hydrodynamics.  In contrast to swimmers, which by definition operate in a fluid environment, gliders are a form of dry active matter \cite{chate2020dry} and their motion does not, in general, depend on hydrodynamic interactions. Instead, gliders experience an adhesive molecular friction, which is a viscous-like frictional drag \cite{tawada1991protein,bormuth2009protein,erbas2012viscous}.

 Our theory takes inspiration from two simple and robust experimental gliders: the biological system of \emph{Mycoplasma mobile} and an artificial system of DNA gliders. Both systems correspond well to the simple glider model we use here, and should be amenable to testing our predictions through future experimental work. \emph{M. mobile} are 450~nm wide and 800~nm long cells \cite{WonjuACSNano2015} (\cref{fig:gliderbkg}a) that use ATPase motors to move protein feet along a surface of sialylated oligosaccharides \cite{MiyataMycoplasma2019} (\cref{fig:gliderbkg}b). Through this gliding mechanism, \emph{M. mobile} move in the direction of their membrane protrusion at speeds of 2.0--4.5~$\mu$m/s. Furthermore \emph{M. mobile} can be modified so that their motion can be externally stopped and started with chemical fuel \cite{UenoyamaPNAS2005} and have trajectory persistence lengths greater than 40~$\mu$m, which is substantially larger than the length of the cell \cite{MorioJBac2016}. These properties make \emph{M. mobile} an ideal biological candidate for trajectory control on the micrometer length scale. Another biological system of note is the  microtubule gliding assay, wherein motor proteins are adhered to a surface and propel microtubule protein filaments across the surface up to speeds of 1 $\mu$m/s \cite{leiblerPRL1995glidingassays,diez2004biomolecular,ray1993kinesin,keya2018dnaswarm}. The trajectories of gliding microtubules are ray-like, and relatively unaffected by thermal motion, having a persistence length of 100-500 $\mu$m \cite{SweetSRep2022}. However, as we predict later, high aspect ratio gliders such as microtubules will exhibit relatively little refraction compared to low aspect ratio shapes such as those with disk or square cross-sections. Recently, Ibusuki and coworkers \cite{IbusukiScience2022} reported a system that is derived from the microtubule gliding assay. The motor proteins were modified to bind particular DNA recognition sequences, rather than microtubule subunits, so that DNA nanotube filaments \cite{Rothemund2004tubes} containing the recognition sequences could glide across surfaces coated with modified motors at speeds of up to 200 nm/s (\cref{fig:gliderbkg}c). As high aspect ratio gliders, we predict that DNA nanotubes will also exhibit relatively little refraction---but in principle, the gliding of DNA nanotubes is not dependent on or enhanced by their shape. Thus it should be possible to create DNA gliders of arbitrary shapes by simply adding recognition sequences to nanostructures made using standard DNA nanotechnology techniques, such as DNA origami \cite{rothemund2006folding} (\cref{fig:gliderbkg}d) or single stranded tiles \cite{SST2012} (\cref{fig:gliderbkg}e). In particular, as we demonstrate later, low aspect ratio shapes should be capable of experiencing strong refraction. 

Encouraged by the properties of {\em M. mobile} and DNA gliders we show, through theory and simulation, that model gliders refract at a resistance discontinuity according to a simple analog of Snell's law. Because Snell's law underlies the fully developed theory of ray optics, intuition and concepts from ray optics become available for the the manipulation of gliders. We demonstrate the power and simplicity of this approach by simulating of gliders interacting with geometrically-patterned friction discontinuities that mimic the behavior of prisms, lenses and traps.  

\section*{Derivation}
\label{Sec:Derivation}
\begin{figure}[H]
    \centering
    \includegraphics[width=.75\textwidth]{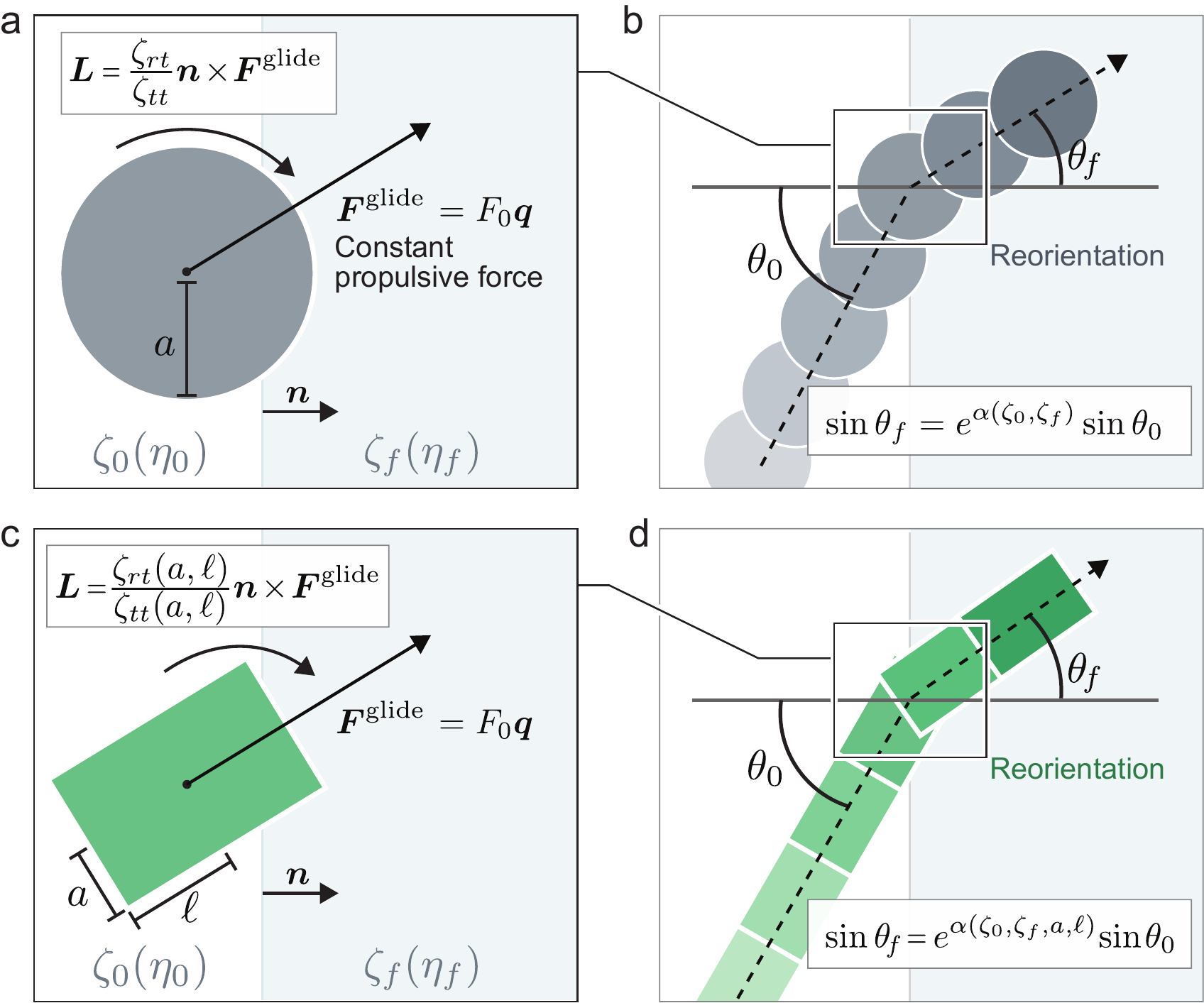}
    \caption{Sketch of discoid (disk-shaped) and rectangular gliders moving across a resistance discontinuity. \textbf{a}, Depiction of the relevant forces and torques on a glider as it travels across a resistance (friction) discontinuity. \textbf{b}, The forces and torques from (\textbf{a}) cause a reorientation of the glider's trajectory around the resistance discontinuity. \textbf{c}, \textbf{d}, Forces and re-orientation of a rectangular glider depend on the aspect ratio of the glider. }
    \label{fig:swim_sketch}
\end{figure}

\vspace*{-0.10in} 
We consider the trajectory of a glider moving across a resistance discontinuity at low Reynold's number. Consistent with our aim to derive a simple theory that can provide intuition for the engineered routing of gliders, we ignore hydrodynamic and Brownian interactions.   
As illustrated in \cref{fig:swim_sketch}, the non-accelerating motion of the glider follows from a balance between the resistance force, $\bF^{resistance}= - \bR_{FU}\!\cdot\!\bU$, and a glide force, $\bF^{glide}$:
\begin{equation}
0 = - \bR_{FU}\!\cdot\!\bU + \bF^{glide} \, .
\label{eq:forcebalance}
\end{equation}
In \cref{eq:forcebalance}, $\bU$ is the translational velocity of the glider, and $\bR_{FU}$ is the resistance function that gives the coupling between the friction  and the velocity.  The resistance tensor depends on the jump in resistivity, and the geometry---the size and shape of the glider as well as its proximity and orientation relative to the discontinuity. 

A glider is an active particle, pushing off of a solid substrate to generate a propulsive glide force of the form:  $\bF^{glide} = F_0 \bq$, where $F_0$ is the magnitude and $\bq$ the direction of propulsion.  To simplify the analysis, we take the resistance tensor to be isotropic and constant, $\bR_{FU} = \zeta_{tt} \bI$, where $\zeta_{tt}$ is the translational resistance coefficient and $\bI$ is the identity tensor. Thus, the velocity is 
\begin{equation}
\bU = F_0 \bq/\zeta_{tt}. 
\label{eq:velocity}
\end{equation}
We consider a constant magnitude glide force $F_0$ and thus the speed of the glider is slower in the region of greater resistivity. For example, for a discoid (disk-shaped) glider (\cref{fig:swim_sketch}a,b) with molecular friction $\eta$ we have a resistance coefficient $\zeta_{tt} \sim \eta a^2$, where $a$ is the glider radius. 

The glide direction $\bq$ changes as a function of time according to
\begin{equation}
\frac{d\bq}{dt} = \bOmega \times \bq \, ,
\label{eq:Omega}
\end{equation}
where $\bOmega$ is the angular velocity of the glider.  The glide angular velocity follows from the torque balance for the force- and torque-free motion
\begin{equation}
0 = - \bR_{L\Omega} \!\cdot\! \bOmega - \bR_{LU}\!\cdot\!\bU \, .
\label{eq:torquebalance}
\end{equation}
In \cref{eq:torquebalance} $\bR_{L\Omega}$ is the resistance tensor coupling the torque ($\bL$) to the angular velocity, and $\bR_{LU}$ couples the torque to the translational velocity.  As for the force-velocity coupling ($FU$), we take the torque-angular velocity coupling to be isotropic and constant: $\bR_{L\Omega} = \zeta_{rr} \bI$; for a discoid glider undergoing rotational molecular friction $\zeta_{rr} \sim \eta a^4$.

The torque-translational velocity ($LU$) coupling arises because as the glider crosses into a region of higher resistivity that portion of the glider engaging the more resistive surface slows down and thus the glider rotates such that its direction of motion tends to align along the normal as illustrated in \cref{fig:swim_sketch}a.  The opposite occurs when moving into a less resistive region.  The $LU$ coupling is a pseudo tensor and since the glider itself is not chiral, it must be of the form $\bR_{LU} = \zeta_{rt} \bepsilon\!\cdot\! \bn$, where $\bepsilon$ is the unit alternating tensor, $\bn$ is the normal to the discontinuity, and $\zeta_{rt}$ is the resistance coefficient. For a disk, the $LU$ coupling only arises if there is a jump in resistivity, $\zeta_{rt} \sim \Delta\eta a^3$.

Combining \cref{eq:forcebalance,eq:torquebalance}, \cref{eq:Omega} becomes
\begin{equation}
\frac{d\bq}{dt} =    \frac{\zeta_{rt}}{\zeta_{rr}}\frac{F_0}{\zeta_{tt}} (\bn \times\bq) \times \bq  =  \frac{\zeta_{rt}}{\zeta_{rr}}\frac{F_0}{\zeta_{tt}} [\bn - \bq (\bq\cdot\bn)] \, .
\label{eq:Omega1}
\end{equation}
Now, $\bn\cdot \bq= \cos\theta$, where $\theta$ is the angle between the normal and the glide direction, and thus \cref{eq:Omega1} gives an equation for the evolution of $\theta(t)$:
\begin{equation}
\frac{d \cos\theta}{\sin^2\theta} =  \frac{\zeta_{rt}}{\zeta_{rr}}\frac{F_0}{\zeta_{tt}}  dt\, .
\label{eq:theta}
\end{equation}
We need to integrate \cref{eq:theta} from the time the glider first touches the discontinuity ($t =0$) with incident angle $\theta_0$ until it fully crosses into the next region at the final time $t_f$, which will then give the out-going angle $\theta_f$.  The time to cross the interface follows from the translational velocity $d\bx/dt = \bU$, and since only the normal component of the velocity is responsible for the glider crossing we have
\begin{equation}
\frac{d (\bn\!\cdot\!\bx)}{dt} = \frac{F_0}{\zeta_{tt}} \bn\!\cdot\! \bq = \frac{F_0}{\zeta_{tt}} \cos\theta \, .
\label{eq:Deltax}
\end{equation}
We can use \cref{eq:Deltax} to replace $dt$ in \cref{eq:theta} to give
\begin{equation}
d x_\perp =  \frac{\zeta_{rr}}{\zeta_{rt}}\frac{\cos\theta \, d \cos\theta}{\sin^2\theta} = - \frac{\zeta_{rr}}{\zeta_{rt}}  d \ln(\sin\theta)\, ,
\label{eq:Deltax1}
\end{equation}
where $x_\perp = \bn\!\cdot\!\bx$ is the amount of the glider that has crossed the interface. For a discoid glider, integrating from 0 to $2a$ relates the initial to the final angle and yields a Snell's law:
\begin{equation}
\sin\theta_f = e^\alpha \sin\theta_0 \, ,
\label{eq:Snell}
\end{equation}
where $\alpha = - 2a \zeta_{rt}/\zeta_{rr}$.

This Snell's law for gliders provides an intuitive principle for how a glider can be reoriented with a resistance discontinuity. The behavior is independent of the magnitude of the propulsive force $F_0$ and the translational resistance $\zeta_{tt}$.  Further by dimensional arguments, the resistance coefficient for $LU$ coupling is proportional to $a^3$ and thus $\alpha$ is independent of the size of the glider.  The validity of this Snell's law and its independence on the glider size are verified by direct simulation below. 

We have made a number of approximations in arriving at this Snell's law. First, we have assumed that the resistance coefficients $\zeta_{rr}$ and $\zeta_{rt}$ are constants  
(Note that $\zeta_{tt}$ cancels out in \cref{eq:Deltax1}). Both coefficients are actually proportional to the local value of the resistance of the surface and thus depend on the portion of the glider in each region. We can include this effect by noting that \cref{eq:Deltax1} can be written as
\begin{equation}
\frac{\zeta_{rt}(x_\perp)}{\zeta_{rr}(x_\perp)} d x_\perp  = -   d \ln(\sin\theta)\, ,
\label{eq:Deltax2}
\end{equation}
and integration from $0$ to $2a$ again recovers Snell's law \cref{eq:Snell} where $\alpha$ is now given by
\begin{equation}
\alpha =  - \int_0^{2a} \frac{\zeta_{rt}(x_\perp)}{\zeta_{rr}(x_\perp)} d x_\perp \, .
\label{eq:alpha}
\end{equation}
However, the exact solution to \cref{eq:alpha} can only be determined through numerical integration. We can approximate $\alpha$ by observing that it is essentially a weighted average of the resistance coefficients in the initial and final states. $\zeta_{rt}$ is proportional to the difference $\Delta \eta = (\eta_f - \eta_0)$ and $\zeta_{rr}$ is proportional to $(\eta_f + \eta_0)/2$. Thus, $\alpha$ becomes

\begin{equation}
\alpha = -  2a  \int_0^1 f(x_\perp; \eta_f/\eta_0) dx_\perp \times \frac{\Delta \eta}{<\eta>} = -C \frac{\Delta \eta}{<\eta>},    
\end{equation}

\noindent where now $x_\perp$ has been made nondimensional by $2a$, and $f$ is a nondimensional function of the distance across the interface $x_\perp$ that depends parametrically on the viscosity ration $\eta_f/\eta_0$. We define $2a$ times the nondimensional integral to be $C(\eta_f/\eta_0)$ giving

\begin{equation}
\alpha = - C\,  \frac{\Delta\eta}{\langle \eta\rangle}\, ,
\label{eq:alpha1}
\end{equation}
where $\Delta\eta = \eta_f - \eta_0$, $ \langle \eta\rangle = (\eta_f + \eta_0)/2$ and the weighting $C$ is an order 1 constant that is weakly dependent of the friction ratio $\eta_f/\eta_0$. Later, we validate our Snell's law for a broad range of friction ratios by using fits (Section~\ref{sectionSI:curvefit}) to micromechanical simulations (Methods) to find $C$ for \cref{eq:alpha1}. 
%(Section~\ref{sectionSI:simulations})

Since the reorientation arises because part of the glider finds itself in a more resistive region, if the glider is very thin relative to its glide axis, then the differential resistance across the body is small and the reorientation should be reduced.  An infinitely thin glider will not reorient at all.  We can account for this shape effect in a simple manner by recognizing that the amount of the glider that has crossed the discontinuity $\Delta x_\perp$ depends on the body shape and the initial orientation $\theta_0$.  For a simple rectangular glider shown in \cref{fig:swim_sketch}c,d, 
\begin{equation}
\Delta x_\perp = \ell \cos\theta_0 + a(1-\cos\theta_0)\, ,
\label{eq:Deltaxell}
\end{equation}
where $\ell$ is the half major length and $a$ is the half minor length.  Using this in \cref{eq:Deltax1} we again have Snell's law, but now  
\begin{equation}
\alpha \sim  - \frac{a}{\ell^2} \left(a + (\ell - a)\cos\theta_0\right) \frac{\Delta\eta}{\langle \eta\rangle} \, ,
\label{eq:alphaell}
\end{equation}
where we have used the geometric scaling that $\zeta_{rt}/\zeta_{rr}  \sim a/ \ell^2$. We have also assumed the instantaneous orientation angle of the body could be approximated with its initial angle $\theta_0$.  When $\ell = a$, \cref{eq:alphaell} reduces to \cref{eq:alpha1}.  The dependence on the aspect ratio is similar to the wavelength dependence of the refractive index. We later use numerical simulations to test this prediction. 

We assumed that any force-angular velocity coupling $\bR_{F\Omega} =  \bR_{LU}^\dag$ was negligible. If $\bR_{F\Omega}$ is included, the force balance \cref{eq:forcebalance} becomes $0 = - \bR_{FU}\!\cdot\!\bU - \bR_{F\Omega}\!\cdot\!\bOmega  + \bF^{glide}$, which, when combined with the angular momentum balance \cref{eq:torquebalance}, will give an angular velocity $\bOmega = - \bR_{L\Omega}^{-1}\!\cdot\!\bR_{LU}\cdot [\bR_{FU} - \bR_{F\Omega}\!\cdot\!\bR_{L\Omega}^{-1}\!\cdot\!\bR_{LU}]^{-1}\cdot \bF^{glide}$.  The additional factor $\bR_{F\Omega}\cdot\bR_{L\Omega}^{-1}\cdot\bR_{LU}$ will add an additional $\theta$ dependence to \cref{eq:theta} and we can only derive a Snell's law under the condition that  $\zeta_{rt}^2/(\zeta_{rr}\zeta_{tt}) \ll 1$. 
  
Furthermore, Non-discoid gliders will not, in general, have isotropic resistance tensors and we can expect, e.g., the force-velocity coupling to have the form $\bR_{FU} = \zeta_\parallel \bq\bq + \zeta_\perp (\bI - \bq\bq)$, where $\zeta_\parallel$ and $\zeta_\perp$ are drag coefficients for motion parallel  and perpendicular  to the glider axis, which we assume be the same as the direction of propulsion $\bq$.  There will also be similar forms for $\bR_{L\Omega}$ and $\bR_{LU}$. Clearly, this complicates the analysis and in general it may be impossible to find a Snell-like analytical expression for the refraction of nondiscoid gliders.

\section*{Analysis of results}
\vspace*{-0.10in} 

\begin{figure}[htb]
    \centering
    \includegraphics[width=.9\textwidth]{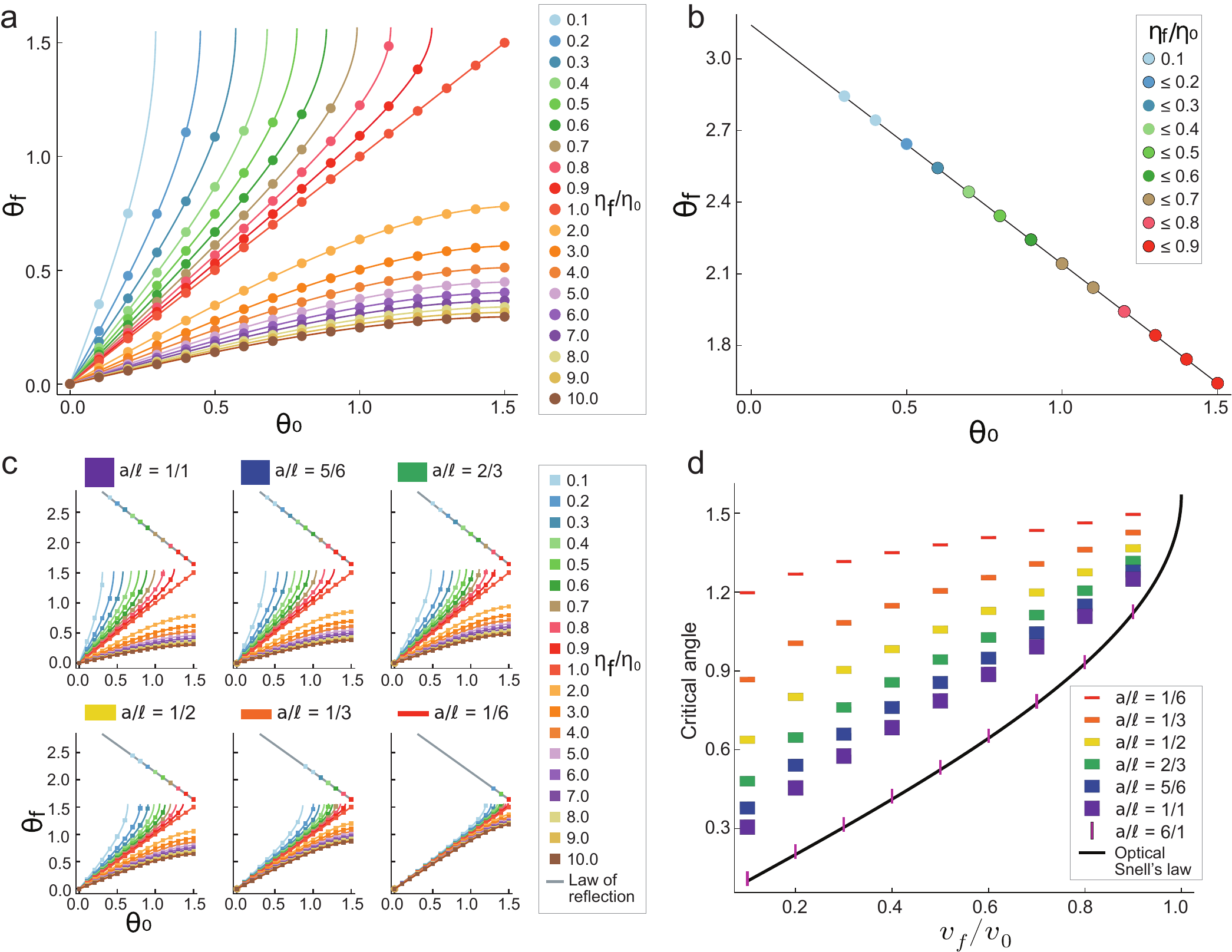}
    \caption{Gliders refract and reflect in a manner analogous to Snell's law. \textbf{a}, Discoid glider refraction as a function of incident angle and friction ratios. Curves represent theory, points represent simulations. \textbf{b}, For incident angles above the critical angle, gliders follow the law of reflection as indicated by the black line. Note that points overlap each other, as indicated in the legend. \textbf{c}, Refraction and reflection for rectangular gliders of varying  thicknesses. Note that points on the law of reflection line overlap each other in the same order as (\textbf{b}). \textbf{d}, Comparison between the critical angles predicted by optical Snell's law and glider Snell's law.}
    \label{fig:snells_law_plots}
\end{figure}

% Discussion of sphere Snell's law 
%Section~\ref{sectionSI:simulations} 
Simulations of the a discoid glider (\cref{fig:snells_law_plots}) closely agree with \cref{eq:Snell,eq:alpha1}; see Methods
for simulation details, Section~\ref{sectionSI:curvefit} for curve fitting details, and \href{https://vimeo.com/605285679}{Video 1} for an example simulation. One prediction from our theory is that refraction is size independent. By simulating gliders of different sizes, we find that the angle of refraction is indeed size invariant (Section \ref{sectionSI:furthervalidation}, \cref{fig:si-validation}a). Further, the form of our Snell's law for a discoid glider predicts that there should be a symmetry about the line $\theta_f = \theta_0$, which we verified by comparing data points and curves across this axis of symmetry (Section \ref{sectionSI:furthervalidation}, \cref{fig:si-validation}b).

As for Snell's law, when $\eta_f/\eta_0 < 1$, \cref{eq:Snell} is valid up to $\theta_f = \frac{\pi}{2}$. For $\theta_f = \frac{\pi}{2}$, the incident critical angle is

\begin{equation}
    \theta_\text{crit} = \arcsin e^{-\alpha}.
    \label{Eq:crit_angle}
\end{equation}

For $\theta_0 > \theta_\text{crit}$ gliders should thus obey the law of reflection,

\begin{equation}
    \theta_f = \pi - \theta_0,
    \label{Eq:law_of_reflection}
\end{equation}
which we confirm with simulations (\cref{fig:snells_law_plots}b). Through symmetry, the critical angle is the same as Snell's window \cite{Lynch2015SnellsWindow}, which is the greatest possible refraction angle for a given $\eta_f/\eta_0 > 1.$

For the rectangular glider, we see that refraction depends on the glider's aspect ratio in \cref{eq:alphaell}. As the glider's aspect ratio becomes smaller relative to its glide axis, the effect of refraction diminishes (\cref{fig:snells_law_plots}c). In the limit that the glider becomes a 1D-line segment along the glide axis, i.e. $a = 0$, the glider undergoes no refraction. In contrast, as the glider's aspect ratio becomes wider relative to its glide axis, there is a greater refraction effect.

Due to the similarities between our theory and optical Snell's law, we compare the two theories in more detail. The ratio of optical refractive indices, $n$, can be rewritten as a ratio of speeds $\frac{n_f}{n_0} = \frac{c/v_f}{c/v_0} = \frac{v_f}{v_0}$, as can the ratio of frictions $\frac{\eta_f}{\eta_0} = \frac{F/v_f}{F/v_0} = \frac{v_f}{v_0}$. Thus, we can compare optical Snell's law, $v_f \sin \theta_f = v_0 \sin \theta_0$, to \cref{eq:Snell,eq:alphaell} by looking at the critical angle as a function of the speed ratio (\cref{fig:snells_law_plots}d). Overall, the critical angles share a similar scaling; gliders tend to refract less than light for the same speed ratio. However, for the wide glider $a/\ell = 6/1$, we find a critical angle and refraction curves (Section~\ref{sectionSI:wide} and \cref{fig:si-wideswim}) that closely follow the optical Snell's law. Based on this finding, we conjecture that as $a/\ell \to \infty$, a rectangular glider will follow the optical Snell's law.  The critical angle for discoid gliders is nearly identical to that for $a/\ell = 1/1$ rectangular gliders and therefore is not plotted in \cref{fig:snells_law_plots}d (compare \cref{fig:si-fitparam}a and \cref{fig:si-fitparam}b, purple dots).

\vspace*{-0.10in}
\section*{Ray optics for gliders}
\vspace*{-0.10in}

\begin{figure}[H]
    \centering
    \includegraphics[width=.98\textwidth]{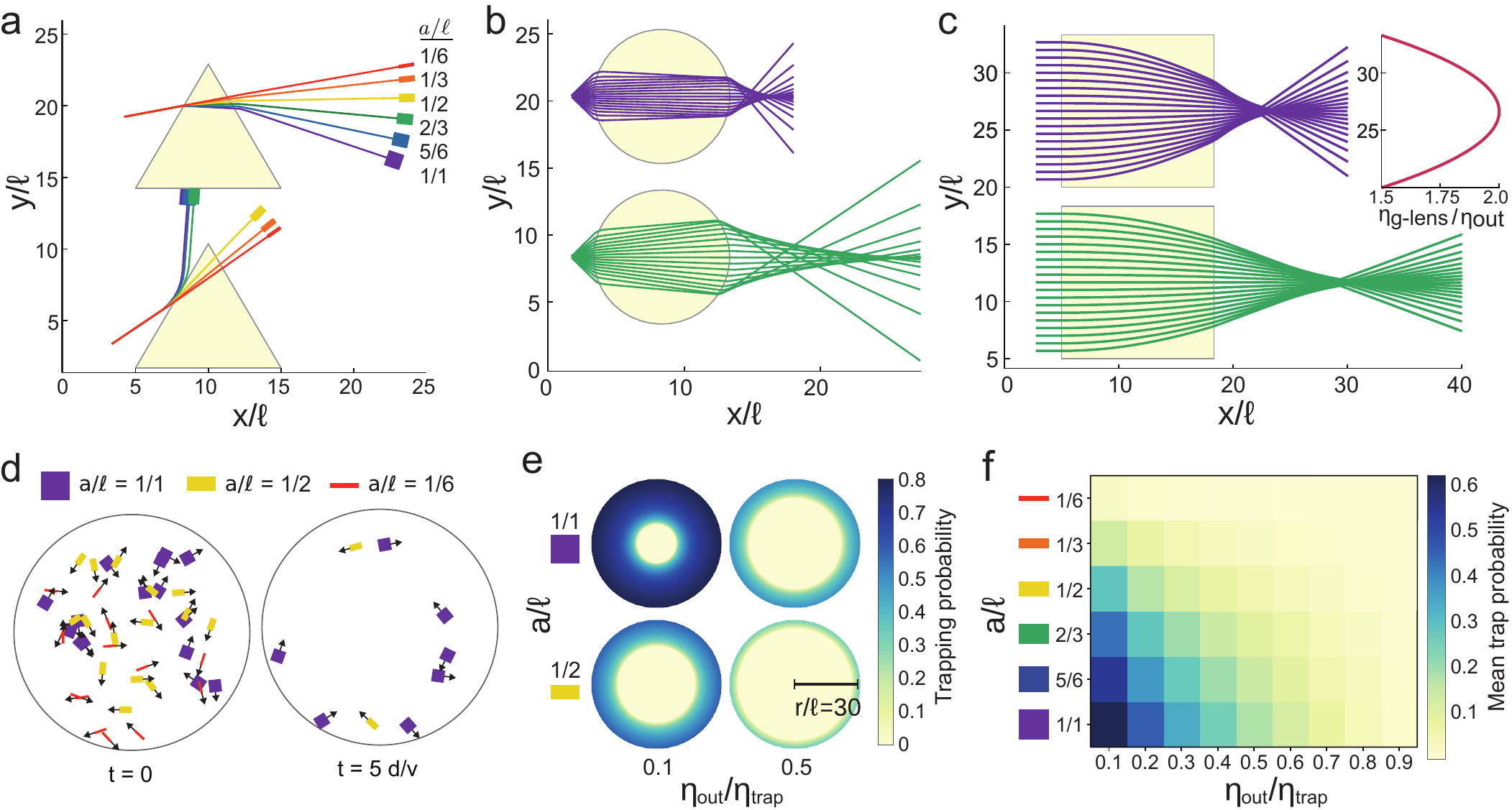}
    \caption{Principles of ray optics can be similarly applied to organize non-interacting gliders. All results are from micromechanical simulations. \textbf{a}, Prism analog disperses rectangular gliders by their shape. Top: Gliders disperse to different angles after passing through a region of higher friction ($\eta_\text{prism}/\eta_\text{out} = 2$). Bottom: A subset of gliders reflect at a discontinuity of lower friction ($\eta_\text{prism}/\eta_\text{out} = 2/3$). \textbf{b}, Ball lens analog has different focal lengths and (spherical aberrations) depending on glider shape. Top: $a/\ell = 1$. Bottom: $a/\ell = 2/3$. \textbf{c}, Gradient lens analog focuses gliders while reducing spherical aberrations. Top:  $a/\ell = 1$, inset shows friction gradient of the lens. Bottom:  $a/\ell = 2/3$. \textbf{d},  Frames of timelapse simulation for a glider trap. Time is measured in $d/v$ where $d$ is the trap diameter and $v$ is the glider speed in the trap. \textbf{e}, Probability of a glider to be trapped given a random initial orientation at the indicated position. \textbf{f}, Mean trapping efficiency increases for smaller friction ratios and decreases for thinner gliders.   }
    \label{fig:razzle-dazzle}
\end{figure}

The simple form of \cref{eq:Snell} suggests an intuitive set of design principles for constructing environments to direct the transport of gliders. Specifically, we consider if the principles of ray optics can also be applied to organize non-interacting gliders. We start by creating a prism, where a triangular region has a friction that differs from the bulk area (\cref{fig:razzle-dazzle}a, \href{https://vimeo.com/659193695}{Video~2}, and \href{https://vimeo.com/605287078}{Video~3}). A polymorphic, multi-shaped, beam of gliders is separated into monomorphic, single-shaped, beams by tuning the friction value of the triangular region. The results from our micromechanical simulations show a finite size effect not present for light. Although refraction and reflection angles are size independent, the centroid trajectory of a glider will appear to bend before encountering the interface (when the leading edge of the glider first contacts the interface) and continue to bend beyond the interface  (until the entire body of the glider is free from the discontinuity) as can be seen in \cref{fig:si-validation}a.

 Having demonstrated that polymorphic beams of gliders can be split into monomorphic beams, we investigate how these monomorphic beams can be shaped. A ball lens that draws gliders toward a focal point can be created using a discoid region with a higher friction than the bulk area (\cref{fig:razzle-dazzle}b). The focus of the lens is a function of the friction ratio, the glider aspect ratio, and the size of the discoid region. Like a ball lens for light, the ball lens for gliders has spherical aberration, i.e. glider trajectories do not all converge at a single focal point. In optics, aberrations are corrected by using compound lenses, aspheric lenses, or lenses with index gradients. We combine refraction with a resistance gradient to create a gradient friction lens (\cref{fig:razzle-dazzle}c) that can focus a collimated beam of monomorphic gliders to a focal point with significantly reduced spherical aberration. Thus, beams of gliders can be sculpted with lenses in a manner similar to light in optics.
 
While the principles of Snell's law can be used to route gliders in free space, they can also be used to confine gliders. We trap gliders in a disk with a higher friction than the bulk area (\cref{fig:razzle-dazzle}d, \href{https://vimeo.com/605286874}{Video~4}). In this design, a glider with an incident angle greater than its critical angle will perpetually reflect off the boundary of the disk with the same incident angle. Gliders whose orientation are below or equal to the critical angle, however, will pass through the trap. Consequently, the trap reaches its steady state by $t= d/v$, where $d$ is the diameter of the disk and $v$ is the glider speed inside the trap. The efficiency of the trap increases for smaller friction ratios (\cref{fig:razzle-dazzle}e,f). Because gliders that are thin relative to their glide axis are trapped much less effectively than thick gliders, discoid traps provide another mechanism to separate gliders based on shape.

\vspace*{-0.10in}
 \section*{Conclusion}
\vspace*{-0.10in} 

Our derivation provides a simple equation for the behavior of gliders at friction discontinuities. It is directly applicable to experimental systems of gliders, such as the \emph{M. mobile} or engineered DNA gliders shown in \cref{fig:gliderbkg}. Further, our force- and torque-balance approach is completely general and pertains to any type of self-propelled particle crossing a friction discontinuity. 

% such as gliding filaments \cite{ray1993kinesin,leiblerPRL1995glidingassays,diez2004biomolecular,keya2018dnaswarm} and mycoplasma \cite{glidingbacteria2013PRL,jarrell2008suprisingwaysprokaryotesmove,MIYATA2008mycoplasmacentipedeinchworm}. Further, our force- and torque-balance approach is completely general and pertains to any type of self-propelled particle crossing a friction discontinuity. %While the addition of terms capturing particle flexibility or hydrodynamics may no longer result in a Snell's law overall, the resistive reorientation we derive will still apply; we suggest that it will be insightful to decompose behavior into that due to resistive reorientation and that due to other effects.

In general, friction discontinuities could be experimentally achieved by patterning a surface with patches of polymer brushes having different lengths or grafting densities \cite{polymerbrush2008}. For DNA gliders in particular, friction discontinuities could created by patterning arrays of DNA tiles \cite{tilings2016} that present DNA extensions whose interaction with the glider could be precisely tuned. Thermal fluctuations or surface defects that cause the glider to reorient will eventually weaken the effects described here, but certain phenomenon should remain, such as the relative efficiency of glider trapping as a function of glider shape. As mentioned in the introduction, trajectory persistence lengths for experimental gliders have been shown to be substantially longer than the dimensions of the glider. These systems therefore possess the right properties for testing the Snell's law we have derived.

% It is therefore reasonable that this phenomenon can be shown experimentally with both the biological and engineered glider systems.

In optics, Snell's law has been derived based on the wave properties of light, or conservation of energy and momentum \cite{NolteFiveSnellsDerivations2018}. Such derivations are inaccessible here, as gliders do not have wave-like properties, momentum is negligible at low Reynold's number, and the energy of a self-propelled glider does not follow traditional energy conservation. Liebchen and L\"{o}wen \cite{Liebchen2019} consider the interesting
problem of determining the optimum navigation strategy for a microswimmer in complex environments, such as those including shear, or vortices. (Other work \cite{routingunderwatergliders2009} treats optimal navigation strategies for macroscopic ocean ``gliders'' and arrives at similar results.) For idealized swimmers that can calculate such optimum strategies and steer accordingly, the authors formulate a variational Fermat's principle that gives an optimal path to minimize travel time,
energy dissipation, or fuel consumption. Given {\em global} knowledge
of the environment this approach leads to geometric paths that follow
Snell's law, or more general refraction, depending on circumstances.
Thus Liebchen and L\"{o}wen's result describes what a
``smart'' swimmer {\em should do}, if it can compute and steer, to
optimize various quantities. Our result, on the other hand, describes what an unguided mindless glider {\em will do} as it
interacts with the environment. Their picture is that {\em forcing}
Snell's law on a swimmer leads to {\em optimal} strategies. Our
picture is that a Snell-like law {\em emerges} from the {\em local}
mechanics; it gives no opportunity to optimize arbitrary
quantities. Their picture connects swimmers to Snell's law and ray
optics by {\em mathematical analogy}---a control algorithm for steering the
swimmer is required to force the mathematics to align, and no analogs to
environmental optical elements can be constructed. Our picture enables
the construction of analogs to optical elements by {\em physical analogy},
with which an unguided glider can interact. In our theory local mechanics, rather
than a global steering algorithm, determines the particle's trajectory. Thus both the
problems and the solutions to our respective work, are only
superficially and coincidentally similar due to their connections
between Snell's law and active particles.

We note that there are a variety of other analogies between Snell's
law and other mechanical systems. For example, the movement of a
particle on a rigid, two-ramp track under gravity yields a mechanical
analog of Snell's law \cite{SnellsRamp2020} when least time is
considered.  This result is similar to that above for swimmer's, in
that it is built into the construction of the system. {\em If} a track is constructed so that the
two ramps of the track have appropriate slopes, {\em then} the
particle will follow a least time path from beginning to the end of
the track. Again, this analogy is forced by the mathematics and
construction of a particular situation, instead of naturally arising
from the physics under any circumstance. 

% Analogies between Snell's law
% and various mechanical situations, like a car axle moving between
% carpet and paper, have been used for decades as pedagogical tools
% \cite{teachingSnell1993}, but to our knowledge our work is the first
% to take such a physical analogy seriously in working out the
% implications and envisioning a full-blown mechanical analog of ray
% optics.

Unlike other derivations of Snell's law, our derivation of Snell's law for gliders relies purely on mechanics, and comes from a transient, shape-dependent torque experienced by the glider during its short crossing of a resistance discontinuity. To us it is surprising that such a local phenomena, so highly unrelated to light, can yield a theory of refraction so similar to that for optics. Future work may expand the theory of Snell's law for gliders to include physical or geometric properties of the glider, such as flexibility or chirality. These properties may allow for more exotic effects, such as a negative index of refraction or an analog of circular polarization.

The framework we describe here readily lends itself to the design of environments that control the transport of gliders. Our numerical simulations have demonstrated that glider ray optics may be possible, so that friction lenses, prisms, and traps might be combined to organize gliders to perform tasks. One possible application of glider ray optics is as an alternative to microfluidic lab-on-a-chip systems, where beams of microscopic gliders could traffic molecular cargo across a chip. A fully autonomous system might be created, in which gliders change shape based on their cargo and are routed accordingly through friction prisms, lenses, and traps.

 %\section*{Methods}
 %\label{methods}
 
 \section*{Methods}
 
 \subsection*{Geometries and Frictions for Ray Optics}
  \label{si:spatialvisc}
\vspace*{-0.10in}

The prism is an equilateral triangle where the length of each side is $10\ell$. For the top prism in \cref{fig:razzle-dazzle}a, the friction ratio is $\eta_\text{prism}/\eta_\text{out} = 2$, while the lower prism is $\eta_\text{prism}/\eta_\text{out} = 2/3$. The ball lens is created by making a discoid region of $\text{radius} / \ell = 5 $ with a friction ratio $\eta_\text{b-lens}/\eta_\text{out} = 8$. The gradient lens is a square with sides of length $\frac{40}{3} \ell$. The parabolic gradient is $\eta_\text{g-lens}/\eta_\text{out} = 2 \left(1 - \left(\frac{ 3\left( y/\ell-20/3 \right)}{40}\right)^2\right)$. The trap is a disk of radius $r = 30 \ell$ with friction $\eta_\text{out}/\eta_\text{trap} = 0.1$.

 \subsection*{Calculation of Trapping Probabilities}
 \label{si:trapprob}
 \vspace*{-0.10in} 
 
Trapping probabilities were determined by calculating the all possible incident angles that a glider could have for every point inside the trap. Trapping occurs when $\theta_0 > \theta_\text{crit}$ because a glider will continuously reflect to have the same incident angle across the circle. The average trapping probability is calculated by integrating the positional trapping probability and dividing it by the area of the trap.
 
 \subsection*{Numerical Simulations}
 \vspace*{-0.10in}
 We verify the theory and model glider ray optics through numerical simulation. We use a position-dependent friction term in the context of a standard Langevin dynamics approach. We solve the equation for the position, $\mathbf r$, of a particle in time 
 \begin{equation}
m\frac{d^2 \mathbf r}{d t^2}=-\gamma(\mathbf r)\frac{d \mathbf r}{d t}+\mathbf F(\mathbf r),
\label{eq:Langevin}
 \end{equation}
 where $m$ is the mass (in practice this value is low relative to the friction, setting a very short inertial timescale but one which nevertheless allows for greater numerical accuracy when iterating forward), $\gamma(\mathbf r)$ is some spatially varying friction, and $\mathbf F$ are the internal and glide forces. 

  The internal forces arise from the microscopic potentials we use in the simulation of which there are two types. Each particle has a {\em steric repulsion}, which gives an upper bound to the possible compressibility of the objects we consider, and for which We use the Weeks-Chandler-Andersen:
 \begin{equation}
 \phi_{\text{WCA}}(r)=
\begin{cases}
4\epsilon\left(\left(\frac{\sigma}{r}\right)^{12}-\left(\frac{\sigma}{r}\right)^6 \right)+\epsilon,& r\le2^{1/6}\sigma\\
0 &r> 2^{1/6}\sigma,
\end{cases}
 \end{equation}
 where $\sigma$ is the hard sphere diameter $\sigma=1$ and $\epsilon=1$.
 
 This is supplemented with the potentials necessary to give our objects {\em structural rigidity}, for which we use harmonic potentials
  \vspace*{-0.15in}
 \begin{equation}
     \phi_{\text{BOND}}(r)=\frac{1}{2}k(r-r_0)^2,
 \end{equation}
 % \vspace*{-0.15in}
 where $r_0$ is some rest length and $k$ is an energy scale. For our simulations we first create an object of a given geometry, and then we add harmonic bonds between nearest neighbor particles where the rest length is taken as the original distance. In general, we set the energy scale $k$ as large as possible in order to ensure structural rigidity of our objects. 
 
 The last element of the simulation is the gliding force exerted on the object. We take this force as acting uniformly on every particle. This force is given by
 \begin{equation}
     \mathbf F^{\text{glide}} = F_0 (\cos\theta,\sin\theta),
 \end{equation}
 where the angle $\theta$ is with respect to the internal axis $\bq$ of the object and the lab frame. We define the internal axes by taking a row of particles within the glider and averaging over the displacements between neighboring particles. The glider's orientation is then updated on each time step. 

\begin{figure}[H]
    \centering
    \includegraphics[width=.98\textwidth]{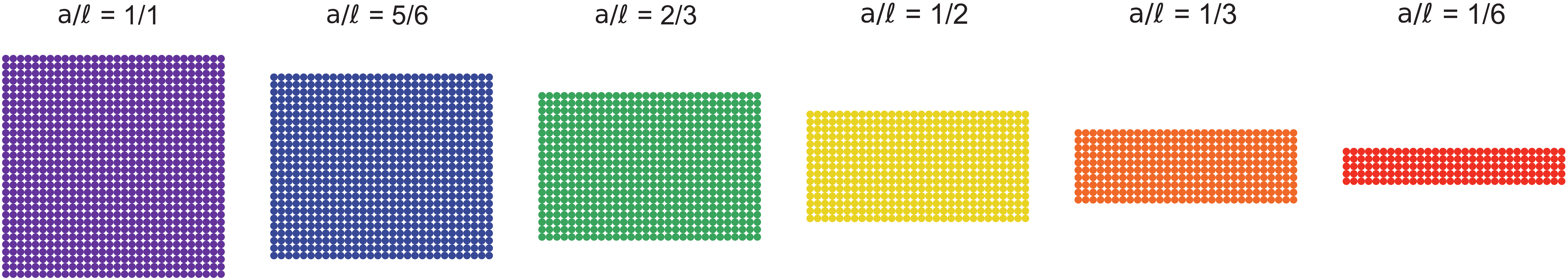}
    \caption{Particle representations of rectangular gliders.}
    \label{fig:si-particles}
\end{figure}

 Our simulation code is written in C++.  Within a glider, the position of each particle is defined by one row of a csv file, where the \textit{x} and \textit{y} positions are the values in the first and second columns, respectfully. Bonds lengths between particles are set by the inter-particle distance of the initial configuration. Discoid gliders have a diameter of 29 particles. Rectangular gliders have a length of 30 particles as show in Fig.~\ref{fig:si-particles}. The spatial dependence of viscosity is determined by the function {\tt spatial\_viscosity} in {\tt mainShape.cpp}. To implement the prism, ball lens, gradient lens and traps, the geometries and viscosities defined were hard-coded in {\tt spatial\_viscosity}.
 
\vspace*{-0.30in} 
 \section*{Acknowledgements}
 \vspace*{-0.10in} 
 
 T.D.R. and P.W.K.R. were funded by the U.S. Department of Energy Award Number Department of Energy award DE-SC0020993. P.W.K.R. was additionally funded by the Office of Naval Research award N00014-18-1-2649. D.O. was partially supported by the U.S. Department of Energy, Office of Science, Office of Basic Energy Sciences, Award Number DE-SC-0010595 and partially through the Moore foundation. J.F.B. was funded by the National Science Foundation Grant 1803662.
 
 \newpage
 \vspace*{-0.70in} 
 \section*{Author contributions}
 \vspace*{-0.10in} 
 
 T.D.R. and P.W.K.R. developed the core concept of the work. J.F.B. and T.D.R. worked on the analytical theory. D.O. built the simulation framework. T.D.R. ran and analyzed the simulations. All authors discussed results and wrote the manuscript.
 
 \vspace*{-0.10in} 
%  \bibliography{zotero.bib}
 \section*{References}
 \vspace*{-0.10in} 
 
 \vspace*{-0.10in} 
 \printbibliography[title={\vspace*{-0.4in}}]

@article{ToyonagaMBio2021,
author = {Takuma Toyonaga  and Takayuki Kato  and Akihiro Kawamoto  and Noriyuki Kodera  and Tasuku Hamaguchi  and Yuhei O. Tahara  and Toshio Ando  and Keiichi Namba  and Makoto Miyata  and Lotte Sogaard-Andersen },
title = {Chained Structure of Dimeric F$_{1}$-like {ATP}ase in Mycoplasma mobile Gliding Machinery},
journal = {mBio},
volume = {12},
number = {4},
pages = {e01414-21},
year = {2021},
doi = {10.1128/mBio.01414-21},
URL = {https://journals.asm.org/doi/abs/10.1128/mBio.01414-21},
eprint = {https://journals.asm.org/doi/pdf/10.1128/mBio.01414-21}
}

@article{IbusukiScience2022,
author = {Ryota Ibusuki  and Tatsuya Morishita  and Akane Furuta  and Shintaro Nakayama  and Maki Yoshio  and Hiroaki Kojima  and Kazuhiro Oiwa  and Ken’ya Furuta },
title = {Programmable molecular transport achieved by engineering protein motors to move on DNA nanotubes},
journal = {Science},
volume = {375},
number = {6585},
pages = {1159-1164},
year = {2022},
doi = {10.1126/science.abj5170},
URL = {https://www.science.org/doi/abs/10.1126/science.abj5170},
eprint = {https://www.science.org/doi/pdf/10.1126/science.abj5170}
}

@article{SweetSRep2022,
  title={Linking path and filament persistence lengths of microtubules gliding over kinesin},
  author={Sweet, May and Kang’iri, Samuel Macharia and Nitta, Takahiro},
  journal={Scientific Reports},
  volume={12},
  number={1},
  pages={1--8},
  year={2022},
  publisher={Nature Publishing Group}
}

@article{HiratsukaPNAS2006,
author = {Yuichi Hiratsuka  and Makoto Miyata  and Tetsuya Tada  and Taro Q. P. Uyeda },
title = {A microrotary motor powered by bacteria},
journal = {Proceedings of the National Academy of Sciences},
volume = {103},
number = {37},
pages = {13618-13623},
year = {2006},
doi = {10.1073/pnas.0604122103},
URL = {https://www.pnas.org/doi/abs/10.1073/pnas.0604122103},
eprint = {https://www.pnas.org/doi/pdf/10.1073/pnas.0604122103}}

@article{MorioJBac2016,
author = {Hanako Morio  and Taishi Kasai  and Makoto Miyata  and W. W. Metcalf },
title = {Gliding Direction of Mycoplasma mobile},
journal = {Journal of Bacteriology},
volume = {198},
number = {2},
pages = {283-290},
year = {2016},
doi = {10.1128/JB.00499-15},
URL = {https://journals.asm.org/doi/abs/10.1128/JB.00499-15},
eprint = {https://journals.asm.org/doi/pdf/10.1128/JB.00499-15}
}

@article{UenoyamaPNAS2005,
author = {Atsuko Uenoyama  and Makoto Miyata },
title = {Gliding ghosts of {\em Mycoplasma mobile}},
journal = {Proceedings of the National Academy of Sciences},
volume = {102},
number = {36},
pages = {12754-12758},
year = {2005},
doi = {10.1073/pnas.0506114102},
URL = {https://www.pnas.org/doi/abs/10.1073/pnas.0506114102},
eprint = {https://www.pnas.org/doi/pdf/10.1073/pnas.0506114102}}

@article{WonjuACSNano2015,
author = {Lee, Wonju and Kinosita, Yoshiaki and Oh, Youngjin and Mikami, Nagisa and Yang, Heejin and Miyata, Makoto and Nishizaka, Takayuki and Kim, Donghyun},
title = {Three-Dimensional Superlocalization Imaging of Gliding Mycoplasma mobile by Extraordinary Light Transmission through Arrayed Nanoholes},
journal = {ACS Nano},
volume = {9},
number = {11},
pages = {10896-10908},
year = {2015},
doi = {10.1021/acsnano.5b03934},
    note ={PMID: 26469129},

URL = { 
        https://doi.org/10.1021/acsnano.5b03934
    
},
eprint = { 
        https://doi.org/10.1021/acsnano.5b03934
    
}

}

@article{erbas2012viscous,
  title={Viscous friction of hydrogen-bonded matter},
  author={Erbas, Aykut and Horinek, Dominik and Netz, Roland R},
  journal={Journal of the American Chemical Society},
  volume={134},
  number={1},
  pages={623--630},
  year={2012},
  publisher={ACS Publications}
}

@article{bormuth2009protein,
  title={Protein friction limits diffusive and directed movements of kinesin motors on microtubules},
  author={Bormuth, Volker and Varga, Vladimir and Howard, Jonathon and Sch{\"a}ffer, Erik},
  journal={Science},
  volume={325},
  number={5942},
  pages={870--873},
  year={2009},
  publisher={American Association for the Advancement of Science}
}

@article{tawada1991protein,
  title={Protein friction exerted by motor enzymes through a weak-binding interaction},
  author={Tawada, Katsuhisa and Sekimoto, Ken},
  journal={Journal of Theoretical Biology},
  volume={150},
  number={2},
  pages={193--200},
  year={1991},
  publisher={Elsevier}
}

@article{stehnach2021viscophobic,
  title={Viscophobic turning dictates microalgae transport in viscosity gradients},
  author={Stehnach, Michael R and Waisbord, Nicolas and Walkama, Derek M and Guasto, Jeffrey S},
  journal={Nature Physics},
  pages={1--5},
  year={2021},
  publisher={Nature Publishing Group}
}

@article{lopez2021dynamics,
  title={Dynamics of a helical swimmer crossing viscosity gradients},
  author={L{\'o}pez, Christian Esparza and Gonzalez-Gutierrez, Jorge and Solorio-Ordaz, Francisco and Lauga, Eric and Zenit, Roberto},
  journal={Physical Review Fluids},
  volume={6},
  number={8},
  pages={083102},
  year={2021},
  publisher={APS}
}

@article{ray1993kinesin,
  title={Kinesin follows the microtubule's protofilament axis.},
  author={Ray, Sanghamitra and Meyh{\"o}fer, Edgar and Milligan, Ronald A and Howard, Jonathon},
  journal={The Journal of Cell Biology},
  volume={121},
  number={5},
  pages={1083--1093},
  year={1993}
}

@article{diez2004biomolecular,
  title={Biomolecular motors operating in engineered environments},
  author={Diez, Stefan and Helenius, Jonne H and Howard, Jonathon},
  journal={Nanobiotechnology: Concepts, Applications and Perspectives},
  volume={1},
  year={2004},
  publisher={John Wiley \& Sons}
}

@article{SnellsRamp2020,
 title = {Mechanical Snell’s Law},
 author = {KyungTae Kim and June-Haak Ee and Kyounghoon Kim and U-Rae Kim and Jungil Lee},
 journal = {Journal of the Korean Physical Society},
 volume = {76},
 number = {4},
 year = {2020},
 pages = {281--285}
}

@article{Liebchen2018,
  doi = {10.1103/physrevlett.120.208002},
  url = {https://doi.org/10.1103/physrevlett.120.208002},
  year = {2018},
  month = may,
  publisher = {American Physical Society ({APS})},
  volume = {120},
  number = {20},
  author = {Benno Liebchen and Paul Monderkamp and Borge ten Hagen and Hartmut L\"{o}wen},
  title = {Viscotaxis
		: Microswimmer Navigation in Viscosity Gradients},
  journal = {Physical Review Letters}
}

@article{Liebchen2019,
  doi = {10.1209/0295-5075/127/34003},
  url = {https://doi.org/10.1209/0295-5075/127/34003},
  year = {2019},
  month = sep,
  publisher = {{IOP} Publishing},
  volume = {127},
  number = {3},
  pages = {34003},
  author = {Benno Liebchen and Hartmut L\"{o}wen},
  title = {Optimal navigation strategies for active particles},
  journal = {{EPL} (Europhysics Letters)}
}

@article{Coppola2021,
  doi = {10.1038/s41598-020-79887-7},
  url = {https://doi.org/10.1038/s41598-020-79887-7},
  year = {2021},
  month = jan,
  publisher = {Springer Science and Business Media {LLC}},
  volume = {11},
  number = {1},
  author = {Simone Coppola and Vasily Kantsler},
  title = {Green algae scatter off sharp viscosity gradients},
  journal = {Scientific Reports}
}

@article{Luo2018,
  doi = {10.1002/adfm.201706100},
  url = {https://doi.org/10.1002/adfm.201706100},
  year = {2018},
  month = apr,
  publisher = {Wiley},
  volume = {28},
  number = {25},
  pages = {1706100},
  author = {Ming Luo and Youzeng Feng and Tingwei Wang and Jianguo Guan},
  title = {Micro-/Nanorobots at Work in Active Drug Delivery},
  journal = {Advanced Functional Materials}
}

@article{Peng2017,
  doi = {10.1039/c6cs00885b},
  url = {https://doi.org/10.1039/c6cs00885b},
  year = {2017},
  publisher = {Royal Society of Chemistry ({RSC})},
  volume = {46},
  number = {17},
  pages = {5289--5310},
  author = {Fei Peng and Yingfeng Tu and Daniela A. Wilson},
  title = {Micro/nanomotors towards in vivo application: cell,  tissue and biofluid},
  journal = {Chemical Society Reviews}
}

@article{Snchez2014,
  doi = {10.1002/anie.201406096},
  url = {https://doi.org/10.1002/anie.201406096},
  year = {2014},
  month = dec,
  publisher = {Wiley},
  volume = {54},
  number = {5},
  pages = {1414--1444},
  author = {Samuel S{\'{a}}nchez and Llu{\'{\i}}s Soler and Jaideep Katuri},
  title = {Chemically Powered Micro- and Nanomotors},
  journal = {Angewandte Chemie International Edition}
}

@article{Ebbens2016,
  doi = {10.1016/j.cocis.2015.10.003},
  url = {https://doi.org/10.1016/j.cocis.2015.10.003},
  year = {2016},
  month = feb,
  publisher = {Elsevier {BV}},
  volume = {21},
  pages = {14--23},
  author = {S.J. Ebbens},
  title = {Active colloids: Progress and challenges towards realising autonomous applications},
  journal = {Current Opinion in Colloid {\&} Interface Science}
}

@article{Li2017,
  doi = {10.1126/scirobotics.aam6431},
  url = {https://doi.org/10.1126/scirobotics.aam6431},
  year = {2017},
  month = mar,
  publisher = {American Association for the Advancement of Science ({AAAS})},
  volume = {2},
  number = {4},
  pages = {eaam6431},
  author = {Jinxing Li and Berta Esteban-Fern{\'{a}}ndez de {\'{A}}vila and Wei Gao and Liangfang Zhang and Joseph Wang},
  title = {Micro/nanorobots for biomedicine: Delivery,  surgery,  sensing,  and detoxification},
  journal = {Science Robotics}
}

@article{Chowdhury2005,
  doi = {10.1016/j.plrev.2005.09.001},
  url = {https://doi.org/10.1016/j.plrev.2005.09.001},
  year = {2005},
  month = dec,
  publisher = {Elsevier {BV}},
  volume = {2},
  number = {4},
  pages = {318--352},
  author = {Debashish Chowdhury and Andreas Schadschneider and Katsuhiro Nishinari},
  title = {Physics of transport and traffic phenomena in biology: from molecular motors and cells to organisms},
  journal = {Physics of Life Reviews}
}

@article{Lauga2009,
  doi = {10.1088/0034-4885/72/9/096601},
  url = {https://doi.org/10.1088/0034-4885/72/9/096601},
  year = {2009},
  month = aug,
  publisher = {{IOP} Publishing},
  volume = {72},
  number = {9},
  pages = {096601},
  author = {Eric Lauga and Thomas R Powers},
  title = {The hydrodynamics of swimming microorganisms},
  journal = {Reports on Progress in Physics}
}

@article{Chou2011,
  doi = {10.1088/0034-4885/74/11/116601},
  url = {https://doi.org/10.1088/0034-4885/74/11/116601},
  year = {2011},
  month = oct,
  publisher = {{IOP} Publishing},
  volume = {74},
  number = {11},
  pages = {116601},
  author = {T Chou and K Mallick and R K P Zia},
  title = {Non-equilibrium statistical mechanics: from a paradigmatic model to biological transport},
  journal = {Reports on Progress in Physics}
}

@article{AbaurreaVelasco_FlexocyteRefraction2019,
	doi = {10.1088/1367-2630/ab5c70},
	url = {https://doi.org/10.1088/1367-2630/ab5c70},
	year = 2019,
	publisher = {{IOP} Publishing},
	volume = {21},
	number = {12},
	pages = {123024},
	author = {Clara Abaurrea-Velasco and Thorsten Auth and Gerhard Gompper},
	title = {Vesicles with internal active filaments: self-organized propulsion controls shape, motility, and dynamical response},
	journal = {New Journal of Physics}
}

@article{theriot2011,
    doi = {10.1371/journal.pbio.1001059},
    author = {Barnhart, Erin L. AND Lee, Kun-Chun AND Keren, Kinneret AND Mogilner, Alex AND Theriot, Julie A.},
    journal = {PLOS Biology},
    publisher = {Public Library of Science},
    title = {An Adhesion-Dependent Switch between Mechanisms That Determine Motile Cell Shape},
    year = {2011},
    volume = {9},
    url = {https://doi.org/10.1371/journal.pbio.1001059},
    pages = {1-19},
    number = {5}
}

@inproceedings{NolteFiveSnellsDerivations2018,
  author = {David D. Nolte},
  booktitle = {Gallileo Unbound: A Path Across Life the Universe and Everything},
  title = {Snell's Law: The Fivefold Way},
  publisher = {Oxford University Press},
  address = {New York},
  pages = {1--10},
  year = {2018}
}

@article {Kantsler1187algae,
	author = {Kantsler, Vasily and Dunkel, J{\"o}rn and Polin, Marco and Goldstein, Raymond E.},
	title = {Ciliary contact interactions dominate surface scattering of swimming eukaryotes},
	volume = {110},
	number = {4},
	pages = {1187--1192},
	year = {2013},
	doi = {10.1073/pnas.1210548110},
	publisher = {National Academy of Sciences},
	issn = {0027-8424},
	URL = {https://www.pnas.org/content/110/4/1187},
	eprint = {https://www.pnas.org/content/110/4/1187.full.pdf},
	journal = {Proceedings of the National Academy of Sciences}
}

@article{keya2018dnaswarm,
author = {Keya, Jakia Jannat and Suzuki, Ryuhei and Kabir, Arif Md. Rashedul and
      Inoue, Daisuke and Asanuma, Hiroyuki and Sada, Kazuki and Hess, Henry and
      Kuzuya, Akinori and Kakugo, Akira},
year = {2018},
title = {{DNA}-assisted swarm control in a biomolecular motor system},
journal = {Nature Communications},
pages = {453},
volume = {9}
}

@article{elfring2019gradientswimmers,
  title = {Active Particles in Viscosity Gradients},
  author = {Datt, Charu and Elfring, Gwynn J.},
  journal = {Phys. Rev. Lett.},
  volume = {123},
  issue = {15},
  pages = {158006},
  numpages = {5},
  year = {2019},
  month = {10},
  publisher = {American Physical Society}
}

@article{rothemund2006folding,
	Author = {Paul W. K. Rothemund},
	Journal = {Nature},
	Number = {7082},
	Pages = {297--302},
	Publisher = {Nature Publishing Group},
	Title = {Folding {DNA} to Create Nanoscale Shapes and Patterns},
	Volume = {440},
	Year = {2006}}

@article{henrichsen1972glidingdef,
author  = {J. Henrichsen},
title = {Bacterial surface translocation: a survey and a classification},
journal = { Bacteriol Rev.},
year =  {1972},
volume = {36},
number = {4},
pages = {478--503},
doi = {doi:10.1128/br.36.4.478-503.1972}
}

@article{leiblerPRL1995glidingassays,
  title = {``Gliding Assays'' for Motor Proteins: A Theoretical Analysis},
  author = {Duke, Thomas and Holy, Timothy E. and Leibler, Stanislas},
  journal = {Phys. Rev. Lett.},
  volume = {74},
  issue = {2},
  pages = {330--333},
  numpages = {0},
  year = {1995},
  month = {1},
  doi = {10.1103/PhysRevLett.74.330},
  url = {https://link.aps.org/doi/10.1103/PhysRevLett.74.330}
}

@article{epifanio2014dissipativeshocksgliding,
author = {Virga, Epifanio G. },
title = {Dissipative shocks behind bacteria gliding},
journal = {Philosophical Transactions of the Royal Society A: Mathematical, Physical and Engineering Sciences},
volume = {372},
number = {2029},
pages = {20130360},
year = {2014},
doi = {10.1098/rsta.2013.0360},
URL = {https://royalsocietypublishing.org/doi/abs/10.1098/rsta.2013.0360},
eprint = {https://royalsocietypublishing.org/doi/pdf/10.1098/rsta.2013.0360}
}

@article{SHROUT2015244,
title = {A fantastic voyage for sliding bacteria},
journal = {Trends in Microbiology},
volume = {23},
number = {5},
pages = {244-246},
year = {2015},
note = {Special Issue: Microbial Translocation},
issn = {0966-842X},
doi = {https://doi.org/10.1016/j.tim.2015.03.001},
url = {https://www.sciencedirect.com/science/article/pii/S0966842X15000554},
author = {Joshua D. Shrout}
}

@article{chate2020dry,
  title={Dry aligning dilute active matter},
  author={Chat{\'e}, Hugues},
  journal={Annual Review of Condensed Matter Physics},
  volume={11},
  pages={189--212},
  year={2020},
  publisher={Annual Reviews}
}

@article{clemmenshess2003chemicaltopography,
author = {Clemmens, John and Hess, Henry and Lipscomb, Ryan and Hanein, Yael and BÃ¶hringer, Karl F. and Matzke, Carolyn M. and Bachand, George D. and Bunker, Bruce C. and Vogel, Viola},
title = {Mechanisms of Microtubule Guiding on Microfabricated Kinesin--Coated Surfaces: Chemical and Topographic Surface Patterns},
journal = {Langmuir},
volume = {19},
number = {26},
pages = {10967--10974},
year = {2003},
doi = {10.1021/la035519y},

URL = { 
        https://doi.org/10.1021/la035519y
    
},
eprint = { 
        https://doi.org/10.1021/la035519y
    
}
}

@article{clemmenshess2004junctions,
author ={Clemmens, John and Hess, Henry and Doot, Robert and Matzke, Carolyn M. and Bachand, George D. and Vogel, Viola},
title  ={Motor-protein ``roundabouts'': Microtubules moving on kinesin-coated tracks through engineered networks},
journal  ={Lab on a Chip},
year  ={2004},
volume  ={4},
issue  ={2},
pages  ={83--86},
publisher  ={The Royal Society of Chemistry},
doi  ={10.1039/B317059D},
url  ={http://dx.doi.org/10.1039/B317059D}
}

@article{Lynch2015SnellsWindow,
author = {David K. Lynch},
journal = {Applied Optics},
number = {4},
pages = {B8--B11},
publisher = {OSA},
title = {Snell's window in wavy water},
volume = {54},
year = {2015},
url = {http://www.osapublishing.org/ao/abstract.cfm?URI=ao-54-4-B8},
doi = {10.1364/AO.54.0000B8},
}

@article{SST2012,
title = {Complex shapes self-assembled from single-stranded DNA tiles},
author = {Wei, Bryan and Dai, Mingjie and Yin, Peng},
journal = {Nature},
year = {2012},
volume = {485},
pages = {623--626}
}

@article{MiyataMycoplasma2019,
title = {Refined Mechanism of Mycoplasma mobile Gliding Based on Structure, {ATP}ase Activity, and Sialic Acid Binding of Machinery},
author = {Miyuki S. Nishikawa and Daisuke Nakane and Takuma Toyonaga and Akihiro Kawamoto and Takayuki Kato and Keiichi Namba and Makoto Miyata },
year = {2019},
volume = {10},
pages = {e02846-19},
journal = {m{B}io}
}

@article{routingunderwatergliders2009,
title = {Routing strategies for underwater gliders},
author = {Russ E. Davis and Naomi E. Leonard and David M. Fratantoni},
journal = {Deep Sea Research {II}},
year = {2009},
pages = {173--187},
volume = {56}
}

@article{Rothemund2004tubes,
author = {Paul W. K. Rothemund and   
Axel Ekani-Nkodo and  Nick Papadakis and  Ashish Kumar and  Deborah Kuchnir Fygenson and Erik Winfree},
title = {Design and Characterization of Programmable DNA Nanotubes},
journal = {Journal of the American Chemical Society},
pages = {16344--16352},
volume = {126},
year = {2004}
}

@article{polymerbrush2008,
title = {Switching of friction by binary polymer brushes},
journal = {Soft Matter}, 
year = {2008}, 
number = {4}, 
pages = {1024--1032},
author = {Mukesh Kumar Vyas and Konrad Schneider and Bhanu Nandan and Manfred Stamm}
}

@article{tilings2016,
title = {A `tile' tale: Hierarchical self-assembly of {DNA} lattices},
author = {Arun Richard Chandrasekarana and Rebecca Zhuo},
year = {2016},
pages = {7--16},
volume = {2},
journal = {Applied Materials Today}
}

@article{metalhelix2010,
author = {Zhang, Li and Peyer, Kathrin E. and Nelson, Bradley J.},
title  = {Artificial bacterial flagella for micromanipulation},
journal  ={Lab on a Chip},
year  ={2010},
volume  ={10},
pages  ={2203--2215}
}

@article{ArtificialChemotaxisLim2014,
author = {Jason S. Park  and Benjamin Rhau  and Aynur Hermann  and Krista A. McNally  and Carmen Zhou  and Delquin Gong  and Orion D. Weiner  and Bruce R. Conklin  and James Onuffer  and Wendell A. Lim },
title = {Synthetic control of mammalian-cell motility by engineering chemotaxis to an orthogonal bioinert chemical signal},
journal = {Proceedings of the National Academy of Sciences},
volume = {111},
number = {16},
pages = {5896--5901},
year = {2014}
}

@article{ImmuneSwarm2022,
  author={Mihlan, Michael and Glaser, Katharina M. and Epple, Maximilian W. and L\"{a}mmermann, Tim},   
  title={Neutrophils: Amoeboid Migration and Swarming Dynamics in Tissues},      
	journal={Frontiers in Cell and Developmental Biology},      
	volume={10},      
	year={2022},      
pages = {871789}
}

@article{MicrotubuleTransportReview2017,
author = {Kari Barlan and Vladimir I. Gelfand},
title = {Microtubule-Based Transport and the Distribution, Tethering, and Organization
of Organelles},
journal = {Cold Spring Harbor Perspectives in Biology},
year = {2017},
volume = {9},
pages = {a025817}
}

@article{BoundariesSteerJanusSpheres2015,
author = {Das, Sambeeta and Garg, Astha and Campbell, Andrew I. and 
      Howse, Jonathan and Sen, Ayusman and Velegol, Darrell and Golestanian, Ramin and Ebbens, Stephen J.},
      year = {2015},
      title = {Boundaries can steer active Janus spheres},
      journal = {Nat. Comm.},
      pages = {8999},
      volume = {6}
}

@article{MolecularRobots2014,
 title = {Molecular Robots with Sensors and Intelligence},
  author = {Hagiya, Masami and  Konagaya, Akihiko and
          Kobayashi, Satoshi and Saito, Hirohide and
          Murata, Satoshi},
  year = {2014},
  journal = {Accounts of Chemical Research},
  pages = {1681--1690},
  volume = {47}
}

@article{TurberfieldRobot2011,
author = {Muscat, Richard A. and Bath, Jonathan and Turberfield, Andrew J.},
title = {A Programmable Molecular Robot},
journal = {Nano Letters},
volume = {11},
number = {3},
pages = {982--987},
year = {2011}
}

@article{QianCargoSort2017,
author = {A.J. Thubagere and W. Li and R.F. Johnson and Z. Chen and S. Doroudi and Y.L. Lee and G. Izatt and S. Wittman and N. Srinivas and D. Woods and E. Winfree and L. Qian},
title = {A cargo-sorting {DNA} robot},
journal = {Science},
year = {2017},
volume = {357},
pages = {6356}
}

@article{ActiveMatterNeedlemanZvonomir2017,
author = {Needleman, Daniel and Dogic, Zvonimir},
year = {2017},
title = {Active matter at the interface between materials science and cell biology},
journal = {Nature Reviews Materials},
pages = {17048},
volume = {2}
}

@article{Fisher2018BioinspiredMicrorobots,
author = {Palagi, Stefano and Fischer, Peer},
year = {2018},
title = {Bioinspired microrobots},
journal = {Nature Reviews Materials},
pages = {113--124},
volume = {3},
}

 %===========================
 % Supplementary materials
 %===========================
 
 \newpage
 
 \setcounter{page}{1}
 \setcounter{section}{0}
 \setcounter{figure}{0}
 \setcounter{equation}{0}
 \renewcommand{\thefigure}{S\arabic{figure}}
 \renewcommand{\theequation}{S\arabic{equation}}
 \renewcommand{\thesection}{S\arabic{section}}
  \renewcommand{\thetable}{S\arabic{table}}
% \setcounter{chapter}{19}
% \chapter*{Supplementary Information}

\noindent {\Huge \bf Supplementary Information}

 \section{Curve Fitting and Determination of the Critical Angle}
 \label{sectionSI:curvefit}
  \vspace*{-0.10in}
 As discussed in the main text, the exact form of the resistance coefficients depends on how the glider straddles the discontinuity. To generate the curves in \cref{fig:snells_law_plots}a,b, and \cref{fig:si-wideswim}, we use a least-squares fit of the simulation data.
 
 \begin{figure}[H]
    \centering
    \includegraphics[width=.98\textwidth]{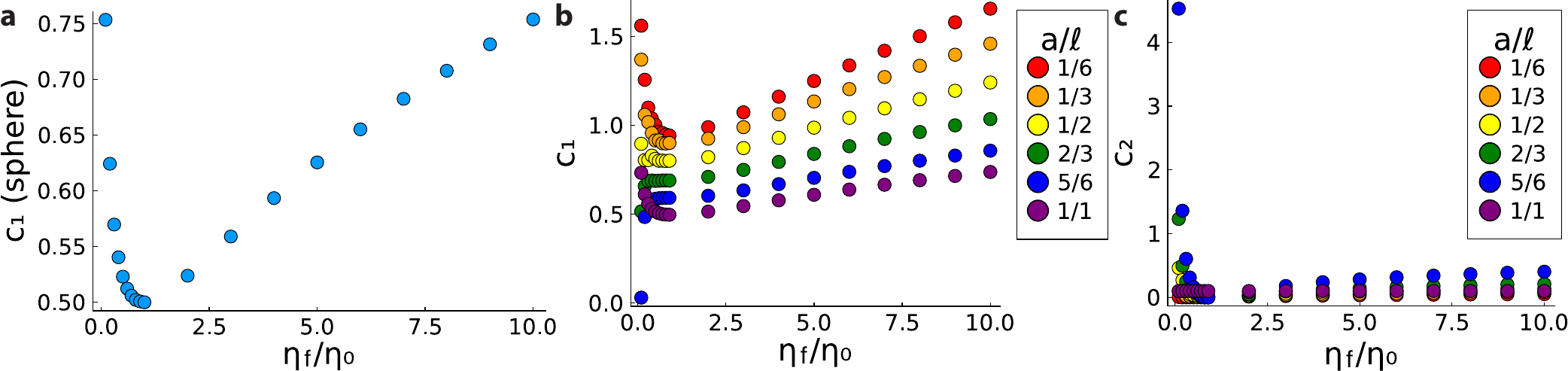}
    \caption{Fit parameters used to generate glider Snell's law curves. \textbf{a}, Fit parameter for a discoid glider. \textbf{b},\textbf{c}, Fit parameters for rectangular glider.}
    \label{fig:si-fitparam}
\end{figure}

 For a discoid glider we use 
 \begin{equation}
 \label{Eq:sphere_fit}
   \sin\theta_f =  \exp\left(-2c_1(\eta_f/\eta_0) \frac{\eta_f - \eta_0}{\eta_f + \eta_0}\right) \sin\theta_0,
 \end{equation}

where $c_1(\eta_f/\eta_0)$ is the free fit parameter that is determined for each friction ratio (\cref{fig:si-fitparam}a). We take the fit for $\sin \theta_f$ to avoid the boundary issue $\arcsin x \in \mathbb{R} , -1 \le x \le 1$. With the values for $c_1(\eta_f/\eta_0)$ in hand, we calculate the critical angle

\begin{equation}
   \theta_\text{crit} = \arcsin \left( \exp\left(2c_1(\eta_f/\eta_0) \frac{\eta_f - \eta_0}{\eta_f + \eta_0}\right) \right).
\end{equation}

For a rectangular glider we follow a similar procedure, but add a second fit parameter since the true $\Delta x_\perp$ is now changed by the glider's rotation as it moves across the interface. We use 

\begin{equation}
\label{Eq:rec_fit}
    \sin \theta_f = \exp\left(-2\frac{a}{\ell^2}\left(c_1(\eta_f/\eta_0,a/\ell) a + c_2(\eta_f/\eta_0,a/\ell) (\ell - a) \cos \theta_0 \right) \frac{\eta_f - \eta_0}{\eta_f + \eta_0} \right) \sin \theta_0,
\end{equation}

where $c_1(\eta_f/\eta_0,a/\ell)$ and $c_2(\eta_f/\eta_0,a/\ell)$ are fit parameters that are functions of friction ratios and the width-to-length ratio of the glider (\cref{fig:si-fitparam}b,c). We then find the rectangular glider's critical angle by numerically solving \cref{Eq:rec_fit} for $\sin \theta_\text{crit} = 1$.

 \section{Further Validation of Snell's Law for Gliders}
 \label{sectionSI:furthervalidation}
 \vspace*{-0.10in}
  \begin{figure}[H]
    \centering
\includegraphics[width=.98\textwidth]{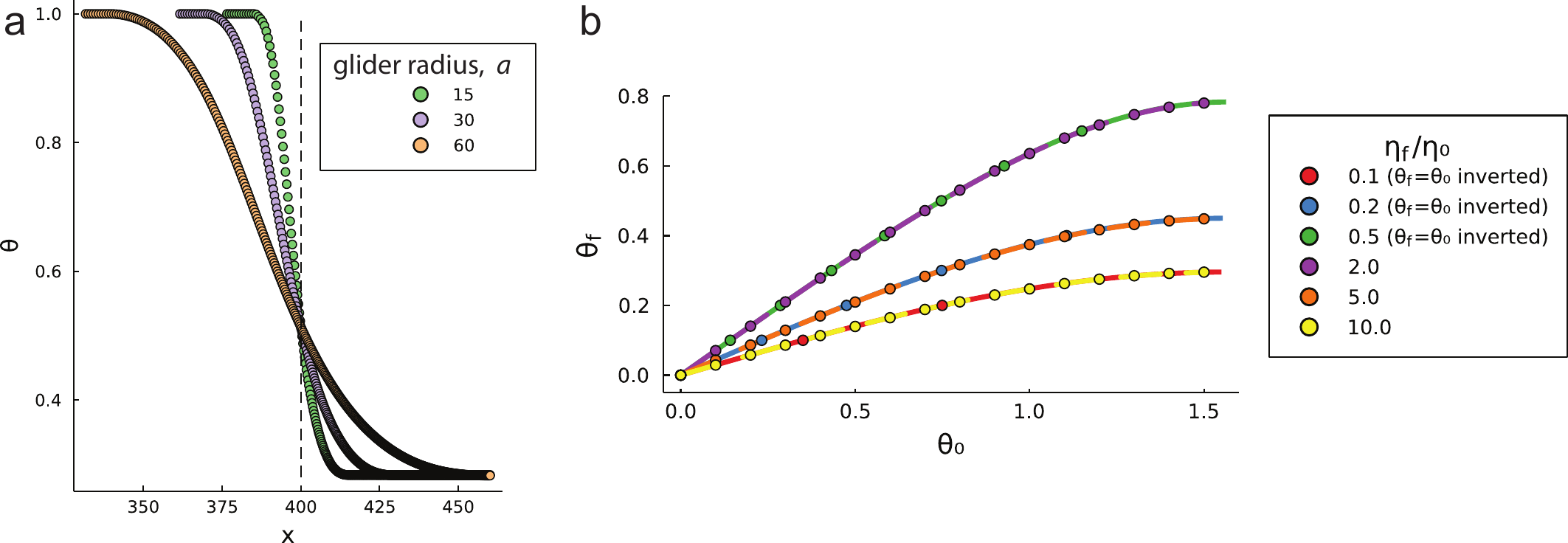}
    \caption{Size independence and symmetry of refraction. \textbf{a}, Comparison of discoid gliders with different radii but the same incident angle ($\theta_0  = 1.0$).  The friction discontinuity is indicated by the vertical dashed line. Glider radii and the $X$-axis are given in terms of the number of particle diameters, for the particles which comprise the gliders. \textbf{b},  Inversion of simulation data and curves for $\eta_f / \eta_0 < 1$ to demonstrate $\theta_f = \theta_0$ symmetry. Data points are simulation and curves are theory. Curves are dashed so that the overlap can be seen.}
    \label{fig:si-validation}
\end{figure}
 
 {\em Size independence.} Given the same incident angle ($\theta_0  = 1.0$), gliders refract to the same final angle ($\theta_f \approx 0.283$), independent of glider radius (\cref{fig:si-validation}a). However, the detailed trajectory of the glider centroid during reorientation differs: larger gliders begin their reorientation earlier and finish their reorientation later than smaller gliders.

 \vspace*{0.10in} 
 
\noindent {\em Symmetry.} \cref{eq:Snell,eq:alpha1} imply that the Snell's law is symmetric across the line $\theta_f = \theta_0$. We test this by taking a subset of simulation data and curves shown in \cref{fig:snells_law_plots}a and inverting those corresponding to $\eta_f/\eta_0 < 1$ across the line $\theta_f = \theta_0$ (\cref{fig:si-validation}b). The tight overlap of the curves and data points to their mirror partner confirms the predicted symmetry.

\section{Wide Rectangular Gliders}
\label{sectionSI:wide}
 \vspace*{-0.10in}
%\vspace*{-0.25in}
  \begin{figure}[H]
    \centering
\includegraphics[width=.98\textwidth]{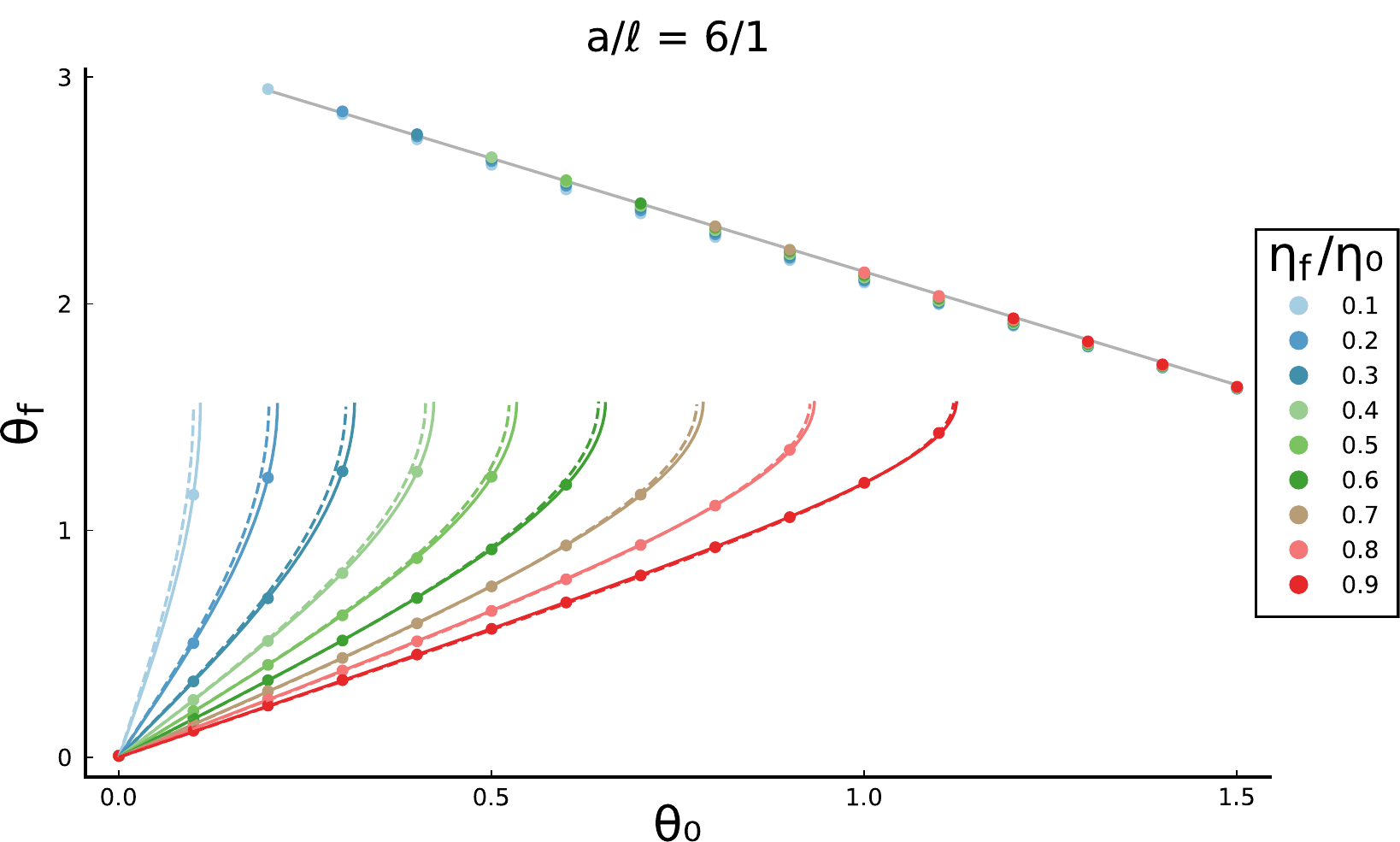}
    \caption{Refraction and reflection for a glider that is wide relative to its glide axis. The optical Snell's law for the equivalent speed ratios is plotted with dashed lines.}
    \label{fig:si-wideswim}
\end{figure}
%\vspace*{-0.15in}

\begin{wraptable}{r}{4cm}
\centering
\begin{tabular}{cc}
\hline
\multicolumn{1}{|l|}{$\eta_f/\eta_0$} & \multicolumn{1}{l|}{\% error} \\ \hline
0.1                             & 12.63                            \\
0.2                             & 6.57                            \\
0.3                             & 4.33                             \\
0.4                             & 2.85                             \\
0.5                             & 1.83                             \\
0.6                             & 1.12                             \\
0.7                             & 0.81                             \\
0.8                             & 0.57                             \\
0.9                             & 0.61                            
\end{tabular}
\caption{Comparison of wide glider curves to optical Snell's law.}
\label{si-table:error}
\end{wraptable}

In \cref{fig:snells_law_plots}c, we establish that refraction becomes less effective as $a/\ell$ becomes smaller. Conversely, as $a/\ell$ becomes larger, the effect of refraction increases. This remains true for gliders with $a/\ell >1$ (\cref{fig:si-wideswim}). In these simulations, the glider's width ($2a$) is 30 particles while the length ($2\ell$) is shortened to 5 particles. For a wide glider, the we must change the geometric scaling $\zeta_{rt}/\zeta_{rr} \sim \ell/a^2$, therefore changing \cref{eq:alphaell} to $\alpha \sim - \frac{l}{a^2}(a + (\ell - a)\cos \theta_0)\frac{\Delta \eta}{<\eta>}$. This change is accounted for when fitting the curves in \cref{fig:si-wideswim}.

We compare this result to the optical Snell's law for both the refraction curves in \cref{fig:si-wideswim} and the critical angles in \cref{fig:snells_law_plots}d using the equivalent speed ratios. We quantify the difference between refraction curves as
\begin{equation}
    \%~\text{error} =\frac{\int \left|\theta_f^\text{glide}(\theta_0)  - \theta_f^\text{light}(\theta_0) \right|d\theta_0}{\int \theta_f^\text{light}(\theta_0) d\theta_0}.
\end{equation}
Based on the low errors (\cref{si-table:error}), we conjecture that in the limit that $a/\ell \to \infty$, the glider will converge to the optical Snell's law.

\end{document}